\documentclass[5p,number,times]{elsarticle}
\usepackage{amsmath,amsfonts,amssymb,color}
\usepackage{graphicx,graphics,booktabs}

\newcommand{\bbR}{\mathbb{R}}
\newcommand{\bbB}{\mathbb{B}}

\newcommand{\bbN}{\mathbb{N}}
\newcommand{\bbP}{\mathbb{P}}
\newcommand{\bbX}{\mathbb{X}}
\newcommand{\bbU}{\mathbb{U}}

\newcommand{\useq}{\mathbf{u}}

\newcommand{\mc}{\mathcal}
\newcommand{\inte}{\operatorname{int}}

\newcommand{\lmax}{\operatorname{\overline{\lambda}}}

\newtheorem{thm}{Theorem}
\newtheorem{rem}[thm]{Remark}
\newtheorem{assum}[thm]{Assumption}
\newtheorem{prop}[thm]{Proposition}
\newproof{pf}{Proof}
\newtheorem{defn}[thm]{Definition}
\newtheorem{lem}[thm]{Lemma}
\newtheorem{cor}[thm]{Corollary}

\makeatletter
\let\mpc@savedenumerate\enumerate
\let\mpc@saved@listi\@listi
\newcommand{\mpc@theoremenumeratemakelabel}[1]{#1\ }
\newcommand{\mpc@thmfixenum}{%
  \setlength{\leftmargini}{5pt}
  \setlength{\labelsep}{0pt}
  \renewcommand{\theenumi}{(\@alph\c@enumi)}
  \renewcommand{\labelenumi}{\theenumi}
  \renewcommand{\enumerate}{
    \renewcommand{\@listi}{\mpc@saved@listi}
    \mpc@savedenumerate
    \renewcommand{\makelabel}{\mpc@theoremenumeratemakelabel}}}

\let\mpc@savedassum\assum
\renewcommand{\assum}{\mpc@savedassum\mpc@thmfixenum}

\allowdisplaybreaks[3]

\begin{document}
\begin{frontmatter}

\title{Linear model predictive control based on \\ polyhedral control Lyapunov functions: theory and applications} 


\author[SG]{Sergio Grammatico}
\author[GP]{Gabriele Pannocchia} 

\address[SG]{Dep. Energy Systems Eng. (DESE), Univ. of Pisa, Pisa, Italy (e-mail: \emph{\texttt{grammatico.sergio@gmail.com}})}
\address[GP]{Dip. Ing. Chim., Chim. Ind. e Sc. Mat. (DICCISM),
  Univ. of Pisa, Pisa, Italy (e-mail: \emph{\texttt{g.pannocchia@diccism.unipi.it}})}

\begin{abstract}
Polyhedral control Lyapunov functions (PCLFs) are exploited in finite-horizon linear model predictive control formulations in order to guarantee the maximal domain of attraction (DoA), in contrast to traditional formulations based on quadratic control Lyapunov functions.
In particular, the terminal region is chosen as the largest DoA, namely the entire controllable set, which is parametrized by a level set of a suitable PCLF. 
Closed-loop stability of the origin is guaranteed either by using an ``inflated'' PCLF as terminal cost or by adding a contraction constraint for the PCLF evaluated at the current state.
Two variants of the formulation based on the inflated PCLF terminal cost are also presented.
In all proposed formulations, the guaranteed DoA is always the entire controllable set, independently of the chosen finite horizon.
Closed-loop inherent robustness with respect to arbitrary, sufficiently small perturbations is also established.
Moreover, all proposed schemes can be formulated as Quadratic Programming problems. 
Numerical examples show the main benefits and achievements of the proposed formulations.
\end{abstract}


\end{frontmatter}
\section{Introduction}

Model predictive control (MPC) algorithms solve a finite-horizon optimal control problem (FHOCP) that includes constraints on states and inputs over the predicted trajectory. 
The first input of the optimal control sequence is injected into the system, and at the successor decision time the FHOCP is solved starting from the new current state.
In order to ensure nominal stability of the origin of the resulting closed-loop system several approaches can be used, e.g. inclusion of a suitable terminal constraint and/or a suitable terminal penalty~\cite[Ch.~2]{rawlings:mayne:2009}, or enforcing the contraction of a suitable control Lyapunov function (CLF) \cite{bemporad:1998, kothare:morari:2000}.
When a terminal constraint is enforced, there is a well defined set of initial states for which the FHOCP is feasible, which is the set of states that can be driven to the terminal region in $N$ steps, where $N$ is the finite horizon. 
Such a set represents the domain of attraction (DoA) of the controller.
The terminal region is often computed assuming (implicitly) that a linear state feedback control law is employed within such region~\cite{gutman:cwikel:1986,michalska:mayne:1993}.
For linear systems, explicit computation of the maximal terminal region is possible~\cite{gilbert:tan:1991}.
A particular case of terminal region is represented by a terminal equality constraint~\cite{mayne:michalska:1990,rawlings:muske:1993}.
The inclusion of a terminal constraint, however, also has some disadvantages, typically associated to the fact that the DoA can be small if a short horizon is used.
In fact, it follows trivially that the DoA can be enlarged by increasing the prediction horizon.
Clearly, longer horizons imply higher computational times, and therefore a trade-off between size of DoA and computational limits is usually necessary.

Terminal penalties are usually employed to take into account (exactly or an upper bound to) the infinite-horizon cost-to-go~\cite{michalska:mayne:1993,rawlings:muske:1993}.
In this way, the optimal value function of FHOCP can be shown to be a Lyapunov function for the closed-loop system, thus implying stability of the origin~\cite{mayne:rawlings:rao:scokaert:2000}.
Moreover, for linear systems with a quadratic cost function, if the terminal penalty is chosen as the solution of the Riccati equation, it is possible to show that the FHOCP yields a solution identical to that of the corresponding infinite-horizon controller~\cite{chmielewski:manousiouthakis:1996,scokaert:rawlings:1998}.
Alternative formulations with neither terminal stabilizing cost nor terminal constraint are also possible (see~\cite[Ch.~6]{grune:pannek:2010} and references therein).

The objective of this paper is to propose linear MPC formulations with the following features:
(i) the DoA is the maximal controllable set irrespectively of the horizon;
(ii) the resulting FHOCP can be posed as a Quadratic Programming (QP) problem;
(iii) for a subset of the DoA the FHOCP yields a solution identical (or similar) to that of the infinite-horizon controller.
To achieve the above goals, we exploit the properties of polyhedral control Lyapunov functions (PCLFs) in the formulation of the FHOCPs.

The use of PCLFs for the constrained stabilization of a linear system traces back to \cite{gutman:cwikel:1986, keerthi:gilbert:1987}. A thorough survey on polyhedral functions for system analysis and control synthesis is \cite{blanchini:1999}. The main advantages of considering PCLF-based stabilization schemes for linear (uncertain) systems are that: 
(i) the maximal (possibly asymmetric) controllable set can be approximated with arbitrary precision; 
(ii) under ``polytopic'' model uncertainties, robust stabilization is equivalent to stabilization by means of a PCLF \cite{blanchini:1994, blanchini:1995}.
From these points of view, polyhedral functions are basically equivalent to composite-quadratic functions \cite{hu:blanchini:2010}. Moreover, constructive algorithms for PCLFs are based on (iterative) linear programming (LP) \cite{blanchini:miani:2008}.

To the best of the authors' knowledge, only a few contributions are available in the literature regarding the use of PCLFs in linear (receding-horizon) MPC formulations. In \cite{bemporad:borrelli:morari:2002, lazar:heemels:weiland:bemporad:pastravanu:2005} infinity norms, namely symmetric polyhedral functions, are employed both in the stage cost and in the terminal cost. On the contrary, \cite{bemporad:1998, kothare:morari:2000} proposed contractive MPC schemes based on quadratic control Lyapunov functions (QCLFs).

The paper is organized as follows. The problem statement is presented in Section~\ref{sec:statement}, together with the basic technical preliminaries.
Several novel MPC formulations based on PCLFs are proposed in Section~\ref{sec:proposed}, while nominal and robust stability analysis is discussed in Section~\ref{sec:stability}. 
Numerical implementations of the proposed MPCs are presented in Section~\ref{sec:numerical}. 
Simulation results are shown in Section~\ref{sec:simulation}. 
The achieved results are summarized in Section~\ref{sec:conclusions}.

\paragraph*{Notation}
Given vectors $x, y$, the inequality $x \leq y$ is intended in a component-wise sense. 
$I$ denotes the identity matrix.
Given a symmetric matrix $A$, the symbols $A \succ 0$ and $A \succeq 0$ mean positive definite and semi-definite, respectively;
moreover, $\lmax_A$ denotes its largest eigenvalue.
$\bbN$ is the set of natural numbers; 
$\bbR$, $\bbR_{>0}$ and $\bbR_{\geq 0}$ denote the sets of real, strictly positive real, and non-negative real numbers, respectively.
$\bbB$ denotes the unitary ball in $\bbR^n$. The interior of a set $S$ is denoted by $\inte(S)$.

\section{Problem statement and technical background} \label{sec:statement}

We consider discrete-time linear time-invariant systems: 
\begin{equation}
\label{eq:linear_system}
x^+ = A x + B u,
\end{equation}
in which $x\in\bbR^n$ and $u\in\bbR^m$ are the state and input at a given time, and $x^+ \in \bbR^n$ is the successor state.
States and inputs are subject to constraints
\begin{equation}\label{eq:constraints}
x(k) \in \bbX \subset \bbR^n , \qquad u(k) \in \bbU \subset \bbR^m  \quad \forall k \in \bbN,
\end{equation}
where $\bbX$ and $\bbU$ are compact and convex polyhedral sets containing the origin. 
In particular, we assume that $\bbX$ contains the origin in its interior, while this is not necessarily required for $\bbU$.
\begin{assum}
The state $x(k)$ is measurable at each sampling time $k\in\bbN$, and the pair $(A,B)$ is stabilizable.
\end{assum}
We use $\useq$ to denote a possibly infinite control sequence $\{ u(k) \mid k \in \bbN \}$, and we use $\phi (k; x, \useq)$ to denote the solution to \eqref{eq:linear_system} if the state at time $0$ is $x$ and the control sequence is $\useq$.
Let $\mc{X}_{\infty} \subseteq \bbX$ be the \textit{maximal controllable set}, defined as
\begin{multline*}
\mc{X}_{\infty} \doteq \{ x \in \bbR^n \mid \exists \useq : u(k) \in \bbU,\; \phi(k; x, \useq) \in \bbX \ \ \forall k \in \bbN, \\
\text{ and } \lim_{k\to\infty}  \phi(k; x, \useq) = 0\}.
\end{multline*}
For any $x \in \mc{X}_{\infty}$, we can define the set of infinite-horizon admissible control sequences as
\begin{multline*}
\mc{U}_{\infty}(x) \doteq \{ \useq \mid u(k) \in \bbU, \; \phi(k; x, \useq) \in \bbX \ \ \forall k \in \bbN, \\\text{ and } \lim_{k\to\infty}  \phi(k; x, \useq) = 0 \}.
\end{multline*}

\begin{assum}
$\mc{X}_{\infty}$ contains the origin in its interior.
\end{assum}

The control objective is the state-feedback stabilization of \eqref{eq:linear_system}, starting from any $x \in \mc{X}_{\infty}$, trying to minimize the quadratic performance cost
\begin{equation*}\label{eq:performance_cost}
V_{\infty} (x,\useq) \doteq \sum_{k=0}^{\infty} \ell (\phi(k; x,\useq),u(k)), 
\end{equation*}
in which $\ell(x,u) \doteq x^\top Q x + u^\top R u$
where $Q \succeq 0 $, $R \succ 0$.
Thus, we consider an infinite-horizon optimal control problem (IHOCP):
\begin{equation} \label{eq:IHOCP}
\mathbb{P}_{\infty}(x): \qquad \min_{\useq} V_{\infty}(x,\useq) \qquad \text{s.t.} \quad \useq \in \mc{U}_{\infty} (x).
\end{equation}
Let $\useq^0(x)$ be the optimal solution of the problem $\mathbb{P}_{\infty}(x)$ and $\kappa(x) \doteq u^0(0;x)$ its first component. Moreover, let $V_\infty^0(x) \doteq V_{\infty}(x,\useq^0(x))$ denote the optimal value of problem $\mathbb{P}_{\infty}(x)$.

\subsection{Preliminaries on polyhedral control Lyapunov functions}

Polyhedral control Lyapunov functions (PCLFs) are particularly suited for the constrained stabilization of \eqref{eq:linear_system}. In fact, in the setting of linear systems subject to polytopic model uncertainties, QCLFs give only sufficient conditions for the robust stabilizability. Conversely, the existence of polyhedral contractive sets and their associated PCLFs is a necessary and sufficient condition for the robust stability of (constrained) uncertain linear systems \cite{blanchini:1999}.
A generic polyhedral function of the second order \cite{hu:blanchini:2010} can be expressed as:
\begin{equation}
\label{pclf}
V_p(x) = \left(\max (F x)\right)^2 \doteq \left(\max_{i \in \mathbb{I}_r} \{F_i x\} \right)^2
\end{equation}
where $\mathbb{I}_r = \{1,2,...,r\}$, 
$F_i \in \bbR^{1 \times n}$ is the $i^{th}$ row of $F$, and the matrix $F$ is such that $\max (F x) \in \bbR_{>0} \; \forall x \in \bbR^n \setminus \{ 0 \}$.
\begin{prop}\label{prop:homogeneous_second_order}
As $V_p(\cdot)$ in \eqref{pclf} is homogeneous of the second order, there exist positive constants $\alpha_1, \alpha_2$ such that:
\begin{equation*}
\alpha_1 \left\|x\right\|^2 \leq V_p(x) \leq \alpha_2 \left\|x\right\|^2 \quad \forall x\in\bbR^n.
\end{equation*}
\end{prop}

It is worth mentioning that, without essential loss of generality, also smoothed PCLFs (via high-order norms) \cite{blanchini:miani:1999} are a universal class of functions for the stabilizability of (constrained) uncertain linear systems. Moreover smoothed PCLFs (unlike standard ones) allow the derivation of explicit formulas for the stabilizing controller \cite{blanchini:miani:1999}.
The fundamental advantage of PCLFs is that the associated polyhedral domain of attraction (DoA) is particularly flexible to cope with control/state constraints since they are capable to approximate the largest
DoA with arbitrary precision \cite{blanchini:1999}. 
Therefore, it can be assumed that the maximal controlled invariant set of \eqref{eq:linear_system} is given by:
\begin{equation}\label{maximal_doa}
\mc{X}_{\infty} = 
\left\{ x \in \bbR^n \mid \ F x  \ \leq \ \underline{1}_{r} \right\},
\end{equation}
where $\underline{1}_{r} \in \bbR^{r}$ is a vector of all ones \footnote{With a slight abuse of notation, by ``maximal controllable set'' $\mc{X}_{\infty}$ we mean that for any given $\epsilon>0$ we can find $F$ such that the size of the controllable set $\mc{X}_{\infty}$ \eqref{maximal_doa} is $\epsilon$-close to the ``true'' maximal controllable set.}.


The computation of the PCLF with the largest controlled DoA can be performed via sequential linear programming (LP), both for discrete-time systems \cite{blanchini:1994} and ``equivalently'' for the class of continuous-time systems \cite{blanchini:1995}.
Note that the procedures in \cite{miani:savorgnan:2005}, \cite{blanchini:miani:2008} compute the maximal (robust) controlled polyhedral set associated to an a-priori fixed \textit{decay rate} $\lambda \in [0,1)$. 
The following Lemma holds true.
\begin{lem}\label{lem:decay}
Let $V_p(\cdot)$ be the PCLF shaping the controlled set $\mathcal{X}_\infty$ \eqref{maximal_doa}. 
Then, for any $x \in \mathcal{X}_\infty$ there exists $u \in \mathbb{U}$ such that:
\begin{equation*}
A x + B u \in \mathcal{X}_\infty, \ V_p(Ax+Bu) \leq \lambda^2 V_p(x), \text{ and } ||u|| \leq c ||x||
\end{equation*}
hold true for some $c \in \bbR_{>0}$.
Moreover, the above condition $V_p(Ax+Bu) \leq \lambda^2 V_p(x)$  is equivalent to 
\begin{equation*}
F (A x + B u) \leq \lambda \max(F x) \underline{1}_{r} .
\end{equation*}
\end{lem}
\begin{pf}
The fact that there exists $u \in \bbU$ such that $A x + B u \in \mathcal{X}_\infty$ and $V_p(Ax+Bu) \leq \lambda^2 V_p(x)$ hold true follows from the definition of CLF. 
To show that such $u$ satisfies $||u|| \leq c ||x||$, we recall that such $u$ is a piecewise-linear function of $x$~\cite{blanchini:1995}, i.e. $u = K^{(h)} x$ where $K^{(h)}$ is the state-dependent gain of a linear controller that guarantees a $\lambda^2$-decay of $V_p(\cdot)$ in a given subset of $\mc{X}_{\infty}$. As the number of different gains $K^{(h)}$ is finite, we define $c \doteq \max_h \{ \| K^{(h)}\| \}$ and obtain that $\|u\| = \|K^{(h)} x\| \leq c \|x\|$.
The last statement follows directly from \eqref{pclf}.
\qed
\end{pf}

\begin{lem}
\label{lemma_alpha3}
There exists a positive constant $\alpha_3$ such that: 
for any $x \in \mc{X}_{\infty}$, there exists $u \in \bbU$ satisfying $Ax + Bu \in \mc{X}_{\infty}$ and
\begin{equation}\label{eq:alpha3}
V_p(A x + B u) - V_p(x) \leq - \alpha_3 \left\| x \right\|^2 .
\end{equation}
\end{lem}
\begin{pf}
From Lemma~\ref{lem:decay}, for any $x \in \mc{X}_{\infty}$, $\exists u \in \bbU$ such that:
\begin{multline*}
V_p (A x + B u) - V_p(x) \leq -(1-\lambda^2) V_p(x) \\
\leq -(1-\lambda^2) \alpha_1 \left\| x \right\|^2 \leq - \alpha_3 \left\| x \right\|^2,
\end{multline*}
holds true for any $0 < \alpha_3 \leq (1-\lambda^2) \alpha_1$. 
\qed
\end{pf}

Given a candidate polyhedral region $\mc{X}$, the minimum admissible decay rate $\lambda$ can be computed by solving \cite{blanchini:1999}:
\begin{multline}\label{eq:lambda-test}
\displaystyle \min_{\lambda, U, W} \lambda \qquad \text{s.t.} \qquad 0 \leq \lambda < 1, \ \ W_{ij} \geq 0 , \\
A X + B U = X W, \ \ \underline{1}_v^\top W = \lambda \underline{1}_v^\top,
\end{multline}
where the $v$ columns of matrix $X \in \mathbb{R}^{n \times v}$ are the $v$ vertices of the given polyhedron $\mc{X}$, $W \in \mathbb{R}^{v \times v}$ and, as a result of the optimization, the columns of $U \in \mathbb{R}^{m \times v}$ are the admissible controls for the vertices of the polyhedron \cite{gutman:cwikel:1986}, \cite{blanchini:1999}.
A solution to \eqref{eq:lambda-test} exists if and only if the given polyhedron is controlled invariant.

\subsection{Basic finite-horizon constrained formulation}

The basic \emph{sub-optimal} solution of the constrained stabilization problem \eqref{eq:IHOCP} is the following finite-horizon constrained formulation. 
Let $\useq$ be a finite-horizon control sequence of length $N$, and let $\phi(k; x, \useq)$ be the corresponding solution to \eqref{eq:linear_system} at time $k$ for the initial state $x(0) = x$.
Define the set of admissible initial states as:
\begin{multline}\label{eq:csetN}
\mc{X}_{N} \doteq\{ x \in \bbR^n \mid \exists \useq : u(k) \in \bbU,\; \\
\phi(k; x, \useq) \in \bbX \ \ \forall k \in \bbN_{0:N-1}, \text{ and }  \phi(N; x, \useq) \in \bbX_f \} ,
\end{multline}
in which $\bbX_f \subseteq \bbX$ is a terminal set later defined.
For any $x \in \mc{X}_{N}$, define the set of finite-horizon admissible control sequences as
\begin{multline}\label{eq:usetN}
\mc{U}_{N}(x) \doteq \{ \useq \mid u(k) \in \bbU, \; \phi(k; x, \useq) \in \bbX \ \ \forall k \in \bbN_{0:N-1}, \\
\text{ and }  \phi(N; x, \useq) \in \bbX_f\} ,
\end{multline}
and the cost
\begin{equation*}
 V_N(x,\useq) \doteq V_f(\phi(N; x,\useq)) + \sum_{k=0}^{N-1} \ell(\phi(k; x,\useq),u(k)) .
\end{equation*}
Consequently, the \emph{basic} finite-horizon optimal control problem (FHOCP) considered is
\begin{equation}\label{standard_rhc}
\mathbb{P}_N(x) : \qquad \min_{\useq} V_N(x,\useq) \qquad \text{s.t.} \quad \useq \in \mc{U}_{N} (x).
\end{equation}

\begin{prop}\label{prop:inclusions}
The following property holds true:
\begin{equation*}
\bbX_f \subseteq \mc{X}_1 \subseteq \cdots \subseteq \mc{X}_N \subseteq \cdots \subseteq \mc{X}_{\infty} \subseteq \bbX.
\end{equation*}
\end{prop}

\begin{rem}
In order to ensure exponential stability of the origin of \eqref{eq:linear_system}, the cost function $V_f(\cdot)$ must satisfy the invariance condition that for any $x\in\bbX_f$, there exists $u\in\bbU$ such that:
\begin{equation*}
 A x + B u \in \bbX_f, \quad \text{and } \quad V_f(A x + B u) - V_f(x)  \leq - \ell(x,u).
\end{equation*}
\end{rem}
For instance, a common choice \cite{chmielewski:manousiouthakis:1996,scokaert:rawlings:1998} for such a function is $V_f(x) \doteq x^\top P x$, where $P \succ 0$ is the (unique) positive definite solution to the discrete-time Algebraic Riccati Equation (ARE)
\begin{equation*}
A^\top P A - P + Q - A^\top P B \left( B^\top P B + R \right)^{-1} B^\top P A = 0.
\end{equation*}
Several options are available to construct $\bbX_f$.
One choice is
\begin{equation}\label{eq:terminal_set}
\bbX_f = \left\{ x \in \bbX \mid \ V_f(x) \leq \alpha \right\} \subseteq \mc{X}_{\infty},
\end{equation}
for some $\alpha \in \bbR_{>0}$, which denotes an ellipsoidal set (possibly maximal), associated to the shape of the Riccati-optimal QCLF $x^\top P x$, such that for any $x \in \bbX_f$, the associated (unconstrained optimal) control is admissible, i.e. $u = K x \in \bbU$, with $K = - \left( B^\top P B + R \right)^{-1} B^\top P A$.
Alternatively \cite{keerthi:gilbert:1988,gilbert:tan:1991}, one can define $\bbX_f$ as the maximal constraint-admissible invariant set for the autonomous system $x^+ = (A + B K) x$, which is generally described by a (possibly large) number of linear inequalities.

\begin{rem}\label{rem:Xf}
When the terminal constraint $\phi(N; x, \useq) \in \bbX_f$ is omitted from the definition of $\mc{U}_N(x)$ \emph{but} the solution to the FHOCP \eqref{standard_rhc}, $\useq^0$, is such that $\phi(N; x, \useq^0) \in \bbX_f$ holds, then it is possible to show that the FHOCP \eqref{standard_rhc} and the IHOCP \eqref{eq:IHOCP} yield the same solution \cite{sznaier:damborg:1987,chmielewski:manousiouthakis:1996,scokaert:rawlings:1998}.
\end{rem}

\begin{rem}
The choice of the Riccati-optimal QCLF is admissible only if $\bbU$ contains the origin in its interior. 
\end{rem}
In fact, if $\bbU$ contains the origin on the boundary, we would have $\bbX_f = \{0\}$.
On the other hand, such assumption is not necessarily required in our problem formulation. 
This further motivates the investigation of (asymmetric) PCLFs and consequently PCLF-based MPC schemes, especially for \emph{short or moderate} control horizons, as discussed later on.
A valid QCLF for the case in which $\bbU$ contains the origin on the boundary is discussed in \cite{rao:rawlings:1999,pannocchia:wright:rawlings:2003}, which yields the \emph{optimal} solution to \eqref{eq:IHOCP} in the limit of \emph{large} control horizon~\cite{pannocchia:wright:rawlings:2003}.

\section{Proposed MPC methods} \label{sec:proposed}

In this section we show how to exploit both the PCLF shaping the maximal controlled DoA and the Riccati-optimal QCLF in the FHOCP. 
In particular, the polyhedral function can be used as a weighted terminal cost or as a guaranteed decay constraint.
We define the set of admissible initial states as:
\begin{multline}\label{eq:csetNp}
\mc{X}^p_{N} \doteq \{ x \in \bbR^n \mid \exists \useq : u(k) \in \bbU,\; \\
\phi(k; x, \useq) \in \bbX \ \ \forall k \in \bbN_{0:N-1}, \text{ and }  \phi(N; x, \useq) \in \mc{X}_{\infty} \} ,
\end{multline}
and, for any $x \in \mc{X}^p_{N}$, the set of admissible control sequences is 
\begin{multline}\label{eq:usetNp}
\mc{U}^p_{N}(x) \doteq \{ \useq \mid u(k) \in \bbU, \; \phi(k; x, \useq) \in \bbX \ \ \forall k \in \bbN_{0:N-1}, \\
\text{ and }  \phi(N; x, \useq) \in \mc{X}_{\infty}\}.
\end{multline}
We first present a number of supporting results.
\begin{lem}
The following property holds true for any $N \in \bbN$:
\begin{equation*}
 \mc{X}_N \subseteq \mc{X}^p_N = \mc{X}_{\infty} .
\end{equation*}
Moreover, for any $x \in \mc{X}_{N}$ there holds:
\begin{equation*}
 \mc{U}_N(x) \subseteq \mc{U}^p_N (x) .
\end{equation*}
\end{lem}
\begin{pf}
From \eqref{eq:csetN}, \eqref{eq:csetNp} and \eqref{eq:usetN}, \eqref{eq:usetNp}, inclusions $\mc{X}_N \subseteq \mc{X}_N^p$ and $\mc{U}_N(x) \subseteq \mc{U}_N^p(x)$ trivially hold because $\bbX_f \subseteq \mc{X}_\infty$ by Proposition \ref{prop:inclusions}. Moreover, $\mc{X}_N^p \subseteq \mc{X}_\infty$ holds as $\mc{X}_\infty$ is the largest controllable set. Thus, $\mc{X}_N^p \supseteq \mc{X}_\infty$ has to be proved for any $N \in \bbN$. Since $\mc{X}_\infty^p \supseteq \cdots \supseteq \mc{X}_2^p \supseteq \mc{X}_1^p \supseteq \mc{X}_0^p$, it is sufficient to show that $\mc{X}_0^p \supseteq \mc{X}_\infty$. From \eqref{eq:csetNp}: $\mc{X}^p_{0} \doteq \{ x \in \bbR^n \mid \phi(0; x, \cdot) = x \in \mc{X}_{\infty} \} = \mc{X}_\infty$.
\qed
\end{pf}

\begin{lem}\label{lem:beta}
For any $x \in \mc{X}_{\infty}$ and $\beta$ satisfying
\begin{equation} \label{eq:beta_asterix}
\beta \geq \beta^* \doteq \frac{\lmax_Q + c^2 \lmax_R}{\alpha_3} ,
\end{equation}
there exists an input $u \in \bbU$ satisfying $(A x + B u) \in \mc{X}_{\infty}$ and 
\begin{equation}\label{eq:beta_costdecay}
\beta {V}_p(A x + Bu) - \beta {V}_p(x) \leq - \ell(x,u) .
\end{equation}
\end{lem}
\begin{pf}
The fact that, for any $x\in\mc{X}_{\infty}$ there exists $u\in\bbU$ such that 
$x^+ = A x + B u \in \mc{X}_{\infty}$ comes the control invariance of $\mc{X}_{\infty}$.
Condition \eqref{eq:beta_costdecay} trivially holds if $x=0$ because we can choose $u=0\in\bbU$ and hence $x^+ = 0$.
In view of Lemma~\ref{lem:decay} and \eqref{eq:alpha3} in Lemma \ref{lemma_alpha3}, 
there always exists $u \in \bbU$ such that:
$\beta {V}_p(x) - \beta {V}_p(A x + B u) \geq \alpha_3 \beta \left\|x\right\|^2 \geq 
(\lmax_Q + c^2 \lmax_R) \left\|x\right\|^2 \geq \ell(x,u)$
holds for any $\beta \geq \beta^* \doteq \frac{\lmax_Q + c^2 \lmax_R}{\alpha_3}$.
\qed
\end{pf}
We next present the two novel PCLF-based formulations, and then discuss two variants of the first one.

\subsection{MPC~1: PCLF-based terminal cost}
As $V_p(\cdot)$ in \eqref{pclf} is a valid PCLF in the whole controllable set $\mc{X}_{\infty}$, 
it can be used as terminal cost.
We introduce a weighting factor $\beta \in \bbR_{>0}$
and define the finite-horizon cost:
\begin{equation*}
\displaystyle V_{N,\beta}(x,\useq) \doteq  \beta V_p (\phi(N; x,\useq)) + \sum_{k=0}^{N-1} \ell(\phi(k; x,\useq),u(k))  ,
\end{equation*}
so that the FHOCP consequently is
\begin{equation}\label{rhc_pclf_2}
\mathbb{P}_N^\beta(x) : \qquad \min_{\useq} V_{N,\beta}(x,\useq) \quad
\text{s.t.} \quad \useq \in \mc{U}^p_{N} (x).
\end{equation}

\begin{rem}
Problem $\mathbb{P}^\beta_N$ \eqref{rhc_pclf_2} is well defined for any $\mc{X}_{\infty}$.
\end{rem}

Let $V_{N,\beta}^0(x)$ denote the optimal value of $\mathbb{P}^\beta_N(x)$. 
As discussed later in Section \ref{section:stability}, if $\beta$ is chosen according to Lemma~\ref{lem:beta}, stability of the origin of the closed-loop system can be proved by showing that $V_{N,\beta}^0(\cdot)$ acts as a Lyapunov function.

\subsection{MPC~2: PCLF-based decay constraint}

Let $\lambda \in [0,1)$ be the guaranteed decay rate of the PCLF (of the first order) $\max( F x )$ 
\cite{blanchini:miani:2008}.
From Lemma~\ref{lem:decay}, for any $x\in\mc{X}_{\infty}$ there exists an admissible control $u \in \bbU$ such that $\max \left( F (A x + B u) \right) \leq \lambda \max (F x)$.
Here we propose the following FHOCP:
\begin{multline}\label{rhc_pclf_1}
\mathbb{P}^\lambda_N(x) : \qquad \min_{\useq} V_N(x,\useq) \quad
\text{s.t.} \quad \useq \in \mc{U}^p_{N} (x), \\
\max (F \phi(1;x,\useq) ) \leq \lambda \max (F x ) ,
\end{multline}
where, we notice that in $V_N(x,\useq)$ we use $V_f(x) = x^\top P x$ with the Riccati-optimal QCLF. The additional (linear) decay constraint $\max (F \phi(1;x,\useq) ) \leq \lambda \max( F x )$ is added to always guarantee a strict decrease of the function $V_p(\cdot)$, that is the one shaping $\mc{X}_{\infty}$ itself, along the closed-loop trajectory. 
\begin{rem}
Problem $\mathbb{P}^\lambda_N$ \eqref{rhc_pclf_1} is well defined for any $x\in\mc{X}_{\infty}$.
\end{rem}
Given any $x \in \mc{X}_{\infty}$, we define the set of admissible inputs as:
\begin{equation*}
 \mc{U}^\lambda_N(x) \doteq \left\{ \useq \in \mc{U}^p_N(x) \mid \max (F \phi(1;x,\useq) ) \leq \lambda \max (F x ) \right\},
\end{equation*}
and, for any $N \in \bbN$, we obviously have that $\mc{U}^\lambda_N(x) \subseteq \mc{U}^p_N(x)$.

\subsection{Two variants of MPC~1 with variable terminal cost}

In view of the knowledge of both the PCLF $V_p(\cdot)$ with maximal controlled DoA and the Riccati-optimal QCLF $V_f(\cdot)$ with IH optimal performance, a dual-mode variant of MPC~1 is here proposed.
We first solve the following problem, which is similar to $\mathbb{P}_N$ in \eqref{standard_rhc}, but has a less stringent terminal constraint $\phi(N; x, \useq) \in \mc{X}_\infty$ instead of $\phi(N; x, \useq) \in \bbX_f$, i.e.:
\begin{equation}
\label{tilde_rhc}
\tilde{\mathbb{P}}_N(x) : \qquad \min_{\useq} V_N(x,\useq) \qquad \text{s.t.} \quad \useq \in \mc{U}_{N}^p (x).
\end{equation}
Let $\tilde{\useq}^0(x)$ be its solution. If $\phi(N; x, \tilde{\useq}^0(x)) \in \bbX_f$, in light of Remark~\ref{rem:Xf}, we have that $\tilde{\useq}^0(x)$ is the optimal solution to the IHOCP.
If, instead, $\phi(N; x, \tilde{\useq}^0(x)) \notin \bbX_f$, we solve problem $\mathbb{P}^\beta_N(x)$ in \eqref{rhc_pclf_2}.
The above dual-mode controller will be referred to as MPC~1a.
\begin{rem}
MPC~1a is well-defined for any $x \in \mc{X}_{\infty}$.
\end{rem}

\begin{rem}
A dual-mode variant is also suited for MPC~2.
\end{rem}

From the formulation $\mathbb{P}_N^\beta$ \eqref{rhc_pclf_2} it is clear that the larger $\beta$, the smaller the effort on minimizing the stage cost $\ell(\cdot)$.
Thus, it may be desirable to use the smallest admissible value for $\beta$.
At each $x \in \mathcal{X}_\infty$, we can compute such smallest admissible $\beta$ as
\begin{equation} \label{eq:beta_star}
\beta^\star(x) \doteq \frac{\bar{\lambda}_Q + c^2 \bar{\lambda}_R}{\alpha_1 (1 - \lambda^\star(x)^2)}.
\end{equation}
where $\lambda^\star(x)$ is the minimal admissible decay rate for the PCLF $V_p(\cdot)$, at given $x$, and it is given by the solution of \eqref{eq:lambda-test}, where $X$, namely $X(x)$, are the vertices of the polyhedron $\{y \in \mathcal{X}_\infty \mid \max(F y) \leq \max( F x ) \}$ having the current state $x$ on its boundary.
Therefore, we consider the cost function 
\begin{equation*}
V_{N,\beta^\star(x)}(x,\useq) \doteq \beta^\star(x) V_p (\phi(N; x,\useq)) + \sum_{k=0}^{N-1} \ell(\phi(k; x,\useq),u(k))  ,
\end{equation*}
and define the FHOCP as
\begin{equation}\label{rhc_pclf_4}
\mathbb{P}_N^{\beta^\star}(x) : \qquad \min_{\useq} V_{N,\beta^\star}(x,\useq) \quad
\text{s.t.} \quad \useq \in \mc{U}^p_{N} (x).
\end{equation}
The above state-dependent terminal cost formulation will be referred to as MPC~1b.

We notice that the weight $\beta^\star(x)$, present in the terminal cost $\beta^\star(x) V_p(\cdot)$ of the cost function $V_{N,\beta^\star(x)}$, decreases as the state $x$ approaches the origin. This means that the optimization problem $\mathbb{P}_N^{\beta^\star}(x)$ \eqref{rhc_pclf_4} weighs more the stage cost as $\left\|x\right\|$ decreases. The result is similar to having a non-homogeneous control Lyapunov function whose shape is close to the one of $V_p(\cdot)$ far from the state-space origin, while close to the locally-optimal one close to the origin \cite{balestrino:caiti:grammatico:2012,balestrino:caiti:grammatico:2012a}.
\begin{rem}
Problem $\mathbb{P}^{\beta^\star}_N$ \eqref{rhc_pclf_4} is well defined for any $\mc{X}_{\infty}$.
\end{rem}

\section{Stability Analysis} \label{sec:stability}
\label{section:stability}

In this section the stability results are provided for the proposed PCLF-based MPC formulations. 
From now on, let $\useq_\beta^0(x)$, $\useq_\lambda^0(x)$, $\tilde{\useq}^0(x)$, $\useq_{\beta^\star}^0(x)$ be the solutions of the problems $\mathbb{P}_N^\beta(x)$ \eqref{rhc_pclf_2}, $\mathbb{P}_N^\lambda(x)$ \eqref{rhc_pclf_1}, $\tilde{\mathbb{P}}_N(x)$ \eqref{tilde_rhc}, $\mathbb{P}_N^{\beta^\star}(x)$ \eqref{rhc_pclf_4} respectively, and let $\kappa_\beta(x) \doteq u_\beta^0(0;x)$, $\kappa_\lambda(x) \doteq u_\lambda^0(0;x)$, $\tilde{\kappa}(x) \doteq \tilde{u}^0(0;x)$, $\kappa_{\beta^\star}(x) \doteq u_{\beta^\star}^0(0;x)$ be their first components, respectively. 

\begin{lem}\label{lem:improving}
For any $\beta$ satisfying condition \eqref{eq:beta_asterix} of Lemma~\ref{lem:beta}
and $N \in \bbN$, the following inequalities hold for any $x \in \mc{X}_\infty$:
\begin{equation*}
V_\infty^0(x) \leq \cdots \leq V_{N+1, \beta}^0(x) \leq V_{N, \beta}^0(x) .
\end{equation*}
\end{lem}
\begin{pf}
We first prove the inequality $V_{N+1, \beta}^0(x) \leq V_{N, \beta}^0(x)$ for any $N\in \bbN$.
Let $\useq_{N,\beta}^0 (x) := (u_{\beta}^0(0;x), u_{\beta}^0(1;x), \ldots, u_{\beta}^0(N-1;x))$ be the optimal solution to problem 
$\mathbb{P}_N^\beta (x)$.
Let $x_{N,\beta}^0 \doteq \phi(N; x, \useq^0_{N,\beta}) \in \mc{X}_{\infty}$.
From Lemma~\ref{lem:beta}, choose any $u^* \in \bbU$ such that $(A x^0_{N,\beta} + B u^*) \in \mc{X}_{\infty}$ and $\beta V_p(A x_{N,\beta}^0 + B u^*) + \ell (x_{N,\beta}^0,u^*) \leq \beta V_{p}(x_{N,\beta}^0)$.
Define the following candidate sequence for problem $\mathbb{P}_{N+1}^\beta (x)$:
$\useq_{N+1,\beta} \doteq (u_{\beta}^0(0;x), u_{\beta}^0(1;x), \ldots, u_{\beta}^0(N-1;x), u^*)$ and notice that
$\useq_{N+1,\beta} \in \mc{U}_{N+1}^p (x)$.
Since $\useq_{N+1,\beta}$ is not necessarily the optimal input sequence for problem $\mathbb{P}_{N+1}^\beta (x)$, we have that:
\begin{multline*}
V_{N+1,\beta}^0 (x) \leq V_{N+1,\beta} (x, \useq_{N+1,\beta}) \doteq \\
V_{N,\beta}^0(x) - \beta V_p(x_{N,\beta}^0) + \ell (x_{N,\beta}^0, u^*) + \beta V_p (A x_{N,\beta}^0 + B u^*) 
\leq V_{N,\beta}^0(x) .
\end{multline*}
To prove $V_{\infty}^0(x) \leq V_{N,\beta}^0(x)$, we note that the sequence $\{V_{N,\beta}^0(x)\}$ is monotonically non-increasing with $N$ and bounded below by 0. 
Thus, it converges to some point $V_{\infty,\beta}^0(x)$. 
We can write:
\allowdisplaybreaks 
\begin{align*}
V_{\infty,\beta}^0(x) & \doteq  \lim_{N\to\infty} V^0_{N,\beta}(x) 
 = \lim_{N\to\infty} \sum_{k=0}^{N-1} \ell (\phi(k; x, \useq_{N,\beta}^0(x)), u_{N,\beta}^0(k;x)) \\
& \quad + \lim_{N\to\infty} \beta V_p (\phi(N; x, \useq_{N,\beta}^0(x)) \\
& \geq \lim_{N\to\infty} \sum_{k=0}^{N-1} \ell (\phi(k; x, \useq_{N,\beta}^0(x)), u_{N,\beta}^0(k;x)) \\
& \geq \sum_{k=0}^{\infty} \ell (\phi(k; x, \useq^0(x)), u^0(k;x)) \doteq V_{\infty}^0 (x) ,
\end{align*}
from which the inequality $V_{\infty}^0 (x) \leq V_{N,\beta}^0(x)$ follows $\forall N\in\bbN$.
\qed
\end{pf}

\begin{cor}\label{cor:quadraticV0}
There exist positive constants $\gamma_1,\gamma_2$ such that:
\begin{equation}\label{eq:gammas}
\gamma_1 \left\| x\right\|^2 \leq V_{N,\beta}^0(x) \leq \gamma_2 \left\| x\right\|^2 .
\end{equation}
\end{cor}
\begin{pf}
From Lemma \ref{lem:improving} we have $V_\infty^0(x) \leq V_{N,\beta}^0(x) \leq V_{0,\beta}^0(x) \doteq \beta V_p(x)$. The optimal cost $x^\top P x$ for the unconstrained system \eqref{eq:linear_system} obviously satisfies $x^\top P x \leq V_\infty^0(x)$. 
Therefore, from Proposition \ref{prop:homogeneous_second_order}, it follows that:
$x^\top P x \leq V_{N,\beta}^0(x) \leq \beta \alpha_2 \left\| x \right\|^2$, 
from which \eqref{eq:gammas} trivially follows for some $\gamma_1, \gamma_2 \in \bbR_{>0}$.
\qed
\end{pf}

\subsection{Nominal stability results}
We use the following notion of exponential stability (ES).
\begin{defn}[Exponential Stability]
Let $\psi(k;x)$ be the solution at time $k$ of the difference equation $x^+ = f(x)$,  
with initial state $x(0) = x$, and let $f(0) = 0$.
The origin of $x^+ = f(x)$ is exponentially stable (ES) on the set $\mc{X} \subseteq \mathbb{R}^n$ if there exist $b \in \bbR_{>0}$ and $\lambda \in (0,1)$ such that for any initial state $x \in \mc{X}$ there holds:
\begin{equation*}
\psi(k;x) \in \mc{X}, \qquad \left\| \psi(k;x) \right\| \leq b \lambda^k \left\| x \right\| \quad \forall k \in \bbN.
\end{equation*}
\end{defn}

\begin{thm}[MPC~1]\label{thm:MPC1}
For any $\beta$ satisfying condition \eqref{eq:beta_asterix} of Lemma~\ref{lem:beta},
the origin of $x^+ = A x + B \kappa_\beta(x)$ is ES on $\mc{X}_{\infty}$.
\end{thm}
\begin{pf}
This result can be proved by applying standard MPC stability results \cite[Thm.~2.24, p.~123]{rawlings:mayne:2009} 
and recalling the results of Proposition~\ref{prop:homogeneous_second_order}, Lemma~\ref{lem:beta} and Corollary~\ref{cor:quadraticV0}.
\qed  
\end{pf}

\begin{thm}[MPC~2]\label{thm:MPC2}
The origin of $x^+ = A x + B \kappa_\lambda(x)$ is ES on $\mc{X}_{\infty}$.
\end{thm}
\begin{pf}
The closed-loop evolution can be written as:
$\psi_\lambda(k+1; x) = A \psi_\lambda(k; x) + B \kappa_\lambda(\psi_\lambda(k; x)) = \phi(1; \psi_\lambda(k; x), \useq_\lambda^0(\psi_\lambda(k; x))$.
Thus, the decay constraint in \eqref{rhc_pclf_1} implies that 
$\psi_\lambda(k; x) \in \mc{X}_\infty$ and 
$\max(F \psi_\lambda(k;x)) \leq \lambda^k \max(F x)$ $\forall k \in  \bbN$.
From Proposition \ref{prop:homogeneous_second_order}, we have
\begin{equation*}
\sqrt{\alpha_1} \left\| \psi_\lambda(k;x) \right\| \leq \max(F \psi_\lambda(k;x)) \leq \lambda^k \max(F x ) \leq \lambda^k \sqrt{\alpha_2} \left\| x \right\| ,
\end{equation*}
which implies
\begin{equation*}
\left\| \psi_{\lambda}(k;x) \right\| \leq \sqrt{\alpha_2 \alpha_1^{-1}} \lambda^k \left\| x \right\| \quad \forall k\in \bbN .
\end{equation*}
Therefore, the origin of $x^+ = A x + B \kappa_\lambda(x)$ is ES on $\mc{X}_{\infty}$.
\qed
\end{pf}

In the proof of next theorem, we consider the set of initial states for which the solution to $\tilde{\bbP}_N(x)$, which has terminal constraint $\phi(N; x, \tilde{\useq}^0) \in \mc{X}_{\infty}$, also satisfies the terminal constraint $\phi(N; x, \tilde{\useq}^0) \in \bbX_f$, i.e.:
\begin{equation}\label{eq:tildeXN}
\tilde{\mc{X}}_{N} \doteq\{ x \in \mc{X}_\infty \mid \phi(N; x, \tilde{\useq}^0) \in \bbX_f \}.
\end{equation}

\begin{rem}
$\bbX_f \subseteq \tilde{\mc{X}}_{N} \subseteq \mc{X}_{N} \subseteq \mc{X}_\infty$ for any $N\in\bbN$.
\end{rem}

\begin{thm}[MPC~1a]\label{thm:MPC3}
Let $\kappa_s(x) \doteq \tilde{\kappa}(x)$ if $\phi(N;x,\tilde{\useq}^0) \in \bbX_f$; 
$\kappa_s(x) \doteq {\kappa}_\beta(x)$, otherwise.
Then, for any $\beta$ satisfying condition \eqref{eq:beta_asterix} of Lemma~\ref{lem:beta},
the origin of $x^+ = A x + B \kappa_s(x)$ is ES on $\mc{X}_{\infty}$.
\end{thm}
\begin{pf}
We observe that if $\psi_s(k;x) \in \tilde{\mc{X}}_{N}$, then $\psi_s(k+1;x) \in \tilde{\mc{X}}_{N}$ because, for any $x \in \tilde{\mc{X}}_{N}$, $\kappa_s(x) = \tilde{u}^0 (x) = \tilde{\kappa}(x)$ is the optimal infinite-horizon control law.
This also implies that $\tilde{\mc{X}}_{N}$ is invariant for the closed-loop system $x^+ = A x + B \kappa_s(x)$, and that the origin of $x^+ = A x + B \kappa_s(x)$ is ES on origin $\tilde{\mc{X}}_{N}$, i.e. we have:
$\|\psi_s(k;x)\| \leq b_1 \lambda_1^k \|x\|$ for all $k\in \bbN$, $x\in\tilde{\mc{X}}_{N}$ and some $b_1 \in \bbR_{>0}$ and 
$\lambda_1 \in (0,1)$.
Assume that the initial state satisfies $x \in \mc{X}_\infty \setminus \tilde{\mc{X}}_{N}$, otherwise the proof is complete.
For all initial states $x \in \mc{X}_\infty \setminus \tilde{\mc{X}}_{N}$, it follows that $\kappa_s(x) = \kappa_\beta(x)$.
Thus, according to Theorem \ref{thm:MPC1}, there exists $\beta^*$ such that, for any $\beta \geq \beta^*$, we have: 
$\left\| \psi_s(k;x) \right\| \leq b_2 \lambda_2^k \left\| x \right\|$ for some $b_2 \in \bbR_{>0}$ and 
$\lambda_2 \in (0,1)$ as long as $\psi_s(k-1;x) \in \mc{X}_\infty \setminus \tilde{\mc{X}}_{N}$.
As $\mc{X}_\infty$ is compact, there exists a finite time $k^* \in \bbN$, $k^* \geq 1$, (dependent on $x$) such that $\psi_s(k^*-1; x) \in \mc{X}_\infty \setminus \tilde{\mc{X}}_{N}$ and $\psi_s(k^*; x) \in \tilde{\mc{X}}_{N}$. 
Furthermore, $\psi_s(k; x) \in \tilde{\mc{X}}_{N}$ for all $k \geq k^*$ because $\tilde{\mc{X}}_{N}$ is invariant for 
$x^+ = A x + B \kappa_s(x)$.
As a consequence, for any $k \leq k^*$ we have:
\begin{equation*}
\left\| \psi_s(k; x) \right\| \leq b_2 \lambda_2^k \|x\| ,
\end{equation*}
whereas for $k > k^*$ we have:
\begin{equation*}
\left\| \psi_s(k; x) \right\| \leq b_1 \lambda_1^{k-k^*} \left\|  \psi_s(k^*; x) \right\| \leq 
b_1 \lambda_1^{k-k^*} b_2 \lambda_2^{k^*} \| x \| .
\end{equation*}
Finally, if we define $\lambda \doteq \max \{\lambda_1, \lambda_2\}$ and $b \doteq \max \{b_1, b_2, b_1 b_2 \}$, it follows that for any $x \in \mc{X}_{\infty}$ the conditions: 
$\psi_s(k;x) \in \mc{X}_{\infty}$ and $\|\psi_s(k;x)\| \leq b \lambda^k \|x\|$ hold for all $k\in\bbN$.
\qed
\end{pf}

\begin{thm}[MPC~1b]\label{thm:MPC1s}
The origin of $x^+ = A x + B \kappa_{\beta^\star}(x)$ is ES on $\mc{X}_{\infty}$.
\end{thm}

\begin{pf}
For any $x \in \mathcal{X}_\infty$ there exists $u \in \bbU$ satisfying $A x + B u \in \ \mc{X}_\infty$, because of the control invariance of $\mc{X}_\infty$, and 
\begin{equation*}
\beta^\star(x) V_p(A x + B u) - \beta^\star(x) V_p(x) \leq -\ell (x,u) ,
\end{equation*}
as $\beta^\star(x) V_p(x) - \beta^\star(x) V_p(A x + B u) \geq \beta^\star(x) \alpha_3(x) ||x||^2 = (\bar{\lambda}_Q + c^2 \bar{\lambda}_R) ||x||^2 \geq \ell(x,u)$ in light of \eqref{eq:beta_star} and $\alpha_3(x) = \alpha_1 (1 - \lambda^\star(x)^2)$.
Then, follow the proof of Theorem \ref{thm:MPC1}. \qed
\end{pf}

\subsection{Inherent robustness}
We briefly discuss further properties of the proposed PCLF-based MPC algorithms about inherent robustness~\cite{grimm:messina:tuna:teel:2004,grimm:messina:tuna:teel:2007}.
In the sake of space, we only focus on MPC~1. 
For robustness analysis we assume that the true system is affected by an unknown bounded disturbance 
$d\in\mathcal{D} \subset \bbR^n$; thus, it evolves as:
\begin{equation*}
x^+ = A x + B u + d .
\end{equation*}
Moreover, we consider the case in which the state is not measured exactly, 
i.e. the measured state is $x_m \doteq x + e$ where $e\in\mathcal{E} \subset \bbR^n$ 
is an unknown bounded noise.
Hence, MPC~1 computes $\useq^0_{\beta} (x_m)$ and implements $\kappa_{\beta}(x_m)$, rather than 
$\kappa_{\beta}(x)$.
The resulting closed loop can be described as a difference inclusion:
\begin{multline}
\label{eq:diff_incl}
 x^+ \in F_{ed} (x) \doteq \{ x^+ \in \bbR^n \mid x^+ = A x + B \kappa_{\beta}(x+e) + d,  \\ 
 \text{ with } d\in\mathcal{D}, e\in\mathcal{E}  \} ,
\end{multline}
because $d$ and $e$ are unknown. 
Let $\psi_{ed}(k; x)$ be \emph{a solution} of the closed-loop system \eqref{eq:diff_incl} for the initial condition $x(0) = x$.
We recall the following definition of Strong Robust Exponential Stability~\cite{pannocchia:rawlings:wright:2011} adapted to the present case.
\begin{defn}
The origin of \eqref{eq:diff_incl} is 
strongly robustly exponentially stable (SRES) on a compact set $\mc{C} \subset \mc{X}_\infty$, 
$0 \in \inte (\mc{C})$, if there exist scalars $b\in\bbR_{>0}$ and $\lambda \in (0,1)$ such that
the following property holds:
given any $\epsilon \in \bbR_{>0}$, there exists $\delta \in \bbR_{>0}$ 
such that for all sequences $\{d(k)\}$ and $\{e(k)\}$ satisfying
\begin{equation}\label{eq:bound}
\|d(k)\| \leq \delta  \text{ and }  \|e(k)\| \leq \delta \quad
\forall k\in\bbN,
\end{equation}
and all $x\in\mc{C}$, it follows that
\begin{subequations}\label{eq:SRES}
\begin{align}
  x_m(k) = x(k)+e(k) \in \mc{X}_\infty , \; x(k) \in \mc{X}_\infty, & \quad \forall k\in\bbN,
  \label{eq:feasible} \\
 \|\psi_{ed}(k; x)\| \leq b \lambda^k \|x\| + \epsilon, &  \quad \forall k\in\bbN . 
 \label{eq:SRES:stab}
\end{align}
 \end{subequations}
\end{defn}
\begin{rem}
In SRES, condition \eqref{eq:feasible} requires that the controller remains feasible at all times for \emph{all} sufficiently small perturbation sequences $\{d(k)\}$ and $\{e(k)\}$.
\end{rem}
We first prove the following useful results.
\begin{prop}\label{prop:V}
$V_{N,\beta}^0 (\cdot)$ is continuous in $\inte (\mc{X}_{\infty})$.
\end{prop}
\begin{pf}
The result follows from \cite[Prop.~12]{grimm:messina:tuna:teel:2004},
\cite[Rem.~27]{pannocchia:rawlings:wright:2011}.
\qed
\end{pf}
\begin{lem}\label{lem:mu}
For any $\mu \in \bbR_{> 0}$, there exists $\delta \in \bbR_{> 0}$ such that:
for all $(x,e,d) \in\mc{X}_{\infty} \times \delta \bbB \times \delta \bbB$ satisfying
$x_m \doteq x + e \in \mc{X}_{\infty}$ and $F_{ed} (x) \subseteq \mc{X}_{\infty}$,
the condition 
\begin{equation}\label{eq:mu}
\max_{x^+ \in F_{ed} (x)} V_{N,\beta}^{0} (x^+) \leq \max\{ \mu, \gamma V_{N,\beta}^{0} (x) \} 
\end{equation}
holds for some $\gamma \in (0,1)$.
\end{lem}
\begin{pf}
From standard stability results \cite[Thm.~2.24, p.~123]{rawlings:mayne:2009} 
and recalling the results of Proposition~\ref{prop:homogeneous_second_order}, Lemma~\ref{lem:beta} and Corollary~\ref{cor:quadraticV0}, we have that $V_{N,\beta}^0(\cdot)$ is an exponential Lyapunov function for the nominal 
closed-loop system $x^+ = A x + B \kappa_{\beta} (x)$. 
Thus, given the measured state $x_m = x + e \in \mc{X}_{\infty}$, the nominal successor state 
$\tilde{x}^+ = A x_m + B \kappa_{\beta}(x_m)$ satisfies:
$V_{N,\beta}^0 (\tilde{x}^+) \leq \bar{\gamma} V_{N,\beta}^0(x_m)$,
for some $\bar{\gamma} \in (0,1)$. 
Choose any $\gamma \in (\bar{\gamma} , 1)$ and define $\rho = \mu (\gamma - \bar{\gamma}) \in \bbR_{>0}$.
We notice that $\tilde{x}^+ - x^+ = A e - d$.
By continuity of $V_{N,\beta}^0(\cdot)$, shown in Proposition~\ref{prop:V}, we can choose 
$\delta_1 \in\bbR_{>0}$ such that
for all $(x,e,d) \in\mc{X}_{\infty} \times \delta_1 \bbB \times \delta_1 \bbB$ satisfying 
$x_m = x + e \in \mc{X}_{\infty}$ and any $x^+\in F_{ed}(x) \subseteq \mc{X}_{\infty}$ there holds:
\begin{equation}\label{eq:delta1}
V_{N,\beta}^0 (x^+) \leq V_{N,\beta}^0 (\tilde{x}^+) + \frac{\rho}{2} .
\end{equation}
By continuity of $V_{N,\beta}^0(\cdot)$, we can also choose $\delta_2 \in\bbR_{>0}$ such that for all 
$(x,e) \in\mc{X}_{\infty} \times \delta_2 \bbB$, the following condition holds:
\begin{equation}\label{eq:delta2}
V_{N,\beta}^0 (\tilde{x}^+) \leq \bar{\gamma} V_{N,\beta}(x) + \frac{\rho}{2} .
\end{equation}
Choose $\delta \doteq \min\{\delta_1, \delta_2\}$;
from \eqref{eq:delta1} and \eqref{eq:delta2}, we obtain:
\begin{equation}\label{eq:delta}
V_{N,\beta}^0 (x^+) \leq \bar{\gamma} V_{N,\beta}(x) + \rho  ,
\end{equation}
for all $(x,e,d) \in\mc{X}_{\infty} \times \delta \bbB \times \delta \bbB$ satisfying $x_m = x + e \in \mc{X}_{\infty}$.
We now define two complementary subsets of $\mc{X}_{\infty}$:
$\bbX_1 \doteq \{ x \in \mc{X}_{\infty} \mid V_{N,\beta}^0(x) \leq \mu \}$ and
$\bbX_2 \doteq \{ x \in \mc{X}_{\infty} \mid V_{N,\beta}^0(x) > \mu \}$, 
and we assume that $\mu$ is not large enough that $\bbX_2$ is empty (otherwise the proof is simpler).
For any $(x,e,d) \in \bbX_1 \times \delta \bbB \times \delta \bbB$ satisfying $x_m = x + e \in \mc{X}_{\infty}$,
and $F_{ed}(x) \subseteq \mc{X}_{\infty}$, it follows from \eqref{eq:delta} that:
$V_{N,\beta}^0 (x^+) \leq \bar{\gamma} \mu + (\gamma - \bar{\gamma}) \mu \leq \mu$.
For any $(x,e,d) \in \bbX_2 \times \delta \bbB \times \delta \bbB$ satisfying $x_m = x + e \in \mc{X}_{\infty}$,
and $F_{ed}(x) \subseteq \mc{X}_{\infty}$, it follows again from \eqref{eq:delta} that:
$V_{N,\beta}^0 (x^+) \leq \bar{\gamma} V_{N,\beta}(x) + (\gamma - \bar{\gamma}) V_{N,\beta}(x) \leq \gamma V_{N,\beta}(x)$.
Thus, for any $(x,e,d) \in \bbX \times \delta \bbB \times \delta \bbB$ satisfying $x_m = x + e \in \mc{X}_{\infty}$,
and $F_{ed}(x) \subseteq \mc{X}_{\infty}$, we have established that \eqref{eq:mu} holds.
\qed
\end{pf}

We now define the set over which SRES is guaranteed.
Consider the largest $\bar{V} \in \bbR_{>0}$ such that for any $\rho \in (0, \bar{V})$
the set $\mc{C}_{\rho} \doteq \{ x \in \mc{X}_{\infty} \mid V_{N,\beta}^0 (x) \leq \bar{V} - \rho \}$ satisfies
$\mc{C}_{\rho} \subset \inte (\mc{X}_{\infty})$.
%
\begin{thm}
The origin of the perturbed closed-loop system \eqref{eq:diff_incl} is SRES on $\mc{C}_{\rho}$.
\end{thm}
\begin{pf}
(\emph{Robust recursive feasibility})
We first prove that given any initial state $x(0)\in\mc{C}_{\rho}$, there exists $\delta\in \bbR_{>0}$ such that \eqref{eq:feasible} holds, i.e. MPC~1 remains feasible at all times for perturbation sequences satisfying \eqref{eq:bound}.
Assume that $x \in \mc{C}_{\rho}$, and choose $\delta_1 \in (0,\rho/2)$. 
Thus, for any $e \in \delta_1 \bbB$, it follows that $x_m = x + e \in \mc{C}_{\rho/2} \subset \inte(\mc{X}_{\infty})$, and MPC~1 is feasible and yields a control $\kappa_{\beta} (x_m)$. 
Then, the nominal successor state is $\tilde{x}^+ = A x_m + B \kappa_{\beta} (x_m)$.
Given that any sub-level set of $V_{N,\beta}^0(\cdot)$ is forward invariant for the nominal closed-loop system, it follows that $\tilde{x}^+ \in \mc{C}_{\rho/2} \subset \inte(\mc{X}_{\infty})$.
Recalling that $\tilde{x}^+ - x^+ = A e - d$, it follows that there exists $\delta_2 > 0$ such that for any $(e,d)\in \delta_2\bbB \times \delta_2\bbB$ the condition $x^+ \in \mc{X}_{\infty}$ holds true.
We can now apply the result of Lemma~\ref{lem:mu}, given any $\mu \leq \bar{V}-\rho$, to obtain that there exists $\delta_3>0$ such that the condition
$V_{N,\beta}^0(x^+) \leq \max\{\mu, \gamma V_{N,\beta}^0(x)\} \leq \bar{V}-\rho$ holds true for any 
$(x,e,d) \in\mc{X}_{\infty} \times \delta_3 \bbB \times \delta_3 \bbB$.
Hence $x^+ \in \mc{C}_{\rho} \subset \inte(\mc{X}_{\infty})$. 
This part of the proof is completed by defining $\delta \doteq \min \{\delta_1,\delta_2,\delta_3\}$.\\
(\emph{Robust exponential stability})
From Corollary~\ref{cor:quadraticV0}, the exist positive constants $\gamma_1,\gamma_2$ such that:
$\gamma_1 \left\| x\right\|^2 \leq V_{N,\beta}^0(x) \leq \gamma_2 \left\| x\right\|^2$.
Choose $\mu \doteq \gamma_1 \epsilon^2$. 
Given a solution $\psi_{ed}(k; x)$ at time $k$, for the initial state $x(0) = x$, 
from Lemma~\ref{lem:mu}, by induction we obtain that for any $(x,e,d) \in\mc{X}_{\infty} \times \delta \bbB \times \delta \bbB$ the condition
\begin{multline*}
\gamma_1 \|\psi_{ed}(k; x)\|^2 \leq V_{N,\beta}^0 (\psi_{ed}(k; x)) \leq \max \{\gamma^k V_{N,\beta}^0(x), \mu \} \\
 \leq \max \{\gamma^k \gamma_2 \|x\|^2, \mu \} \leq \max \{\gamma^k \gamma_2 \|x\|^2, \gamma_1 \epsilon^2 \}
\end{multline*}
holds true (if necessary we can reduce $\delta$).
This implies that:
\[
\|\psi_{ed}(k; x)\| \leq \max \{ b \gamma^k \|x\|, \epsilon \} \leq b \gamma^k \|x\| + \epsilon ,
\]
for $b \doteq \sqrt{\gamma_2/\gamma_1}$ and $\lambda \doteq \sqrt{\gamma} \in (0,1)$. 
\qed
\end{pf}

\section{Numerical implementations} \label{sec:numerical}

Problems \eqref{rhc_pclf_2} and \eqref{rhc_pclf_1} are not posed as standard QP problems. 
However, we have the following results.
\begin{prop} \label{prop:linear_equivalence}
For any $\xi \in \bbR_{\geq 0}$
\begin{equation*}
\left\{x \in \bbR^n \mid V_p(x) \leq \xi^2 \right\}  = \left\{ x \in \bbR^n \mid F x \leq \xi \underline{1}_{r} \right\}.
\end{equation*}
\end{prop}
\begin{pf}
\begin{multline*}
V_p(x) \doteq (\max(Fx))^2 = \left(\max_{i \in \mathbb{I}_r} F_i x \right)^2 \leq \xi^2 \Leftrightarrow \\
\max_{i \in \mathbb{I}_r}\{F_i x\} \leq \xi \Leftrightarrow F_i x \leq \xi \ \ \forall i \in \mathbb{I}_r \Leftrightarrow F x \leq \xi \underline{1}_{r}. \quad
\text{\qed}
\end{multline*}
\end{pf}

\begin{prop}\label{prop:QPbeta}
The nonlinear optimization problem $\mathbb{P}^\beta_N$ \eqref{rhc_pclf_2} is equivalent to the QP problem
\begin{multline}\label{rhc_pclf_2_qp}
\bar{\mathbb{P}}^\beta_N(x) : \qquad \min_{\useq, \xi} \bar{V}_{N,\beta}(x,\useq,\xi) \quad
\text{s.t.} \quad \useq \in \mc{U}^p_{N} (x), \ \ \xi \in [0,1] \\
F \phi(N; x, \useq) \leq \xi \underline{1}_{r},
\end{multline}
having cost function
\begin{equation}\label{eq:cost_function_xi}
\bar{V}_{N,\beta}(x,\useq,\xi) = \beta \xi^2 + \sum_{k=0}^{N-1}\{ \ell(x(k),u(k)) \}.
\end{equation}
\end{prop}
\begin{pf}
Problem $\mathbb{P}^\beta_N$ \eqref{rhc_pclf_2} can be reformulated as
\begin{equation*}
\min_{\useq,\xi} \bar{V}_{N,\beta}(x,\useq,\xi) \quad \text{s.t.} \quad \useq \in \mc{U}^p_{N} (x), \ V_p(\phi(N; x, \useq)) \leq \xi^2
\end{equation*}
with cost function $\bar{V}_{N,\beta}(x,\useq,\xi)$ in \eqref{eq:cost_function_xi}.
The QP formulation~\eqref{rhc_pclf_2_qp} is finally recovered in view of Proposition \ref{prop:linear_equivalence}.
\qed
\end{pf}

\begin{rem}
In Proposition~\ref{prop:QPbeta}, ``equivalent'' is intended in the following sense: 
the optimal values of \ $\mathbb{P}^\beta_N$ \eqref{rhc_pclf_2} and $\bar{\mathbb{P}}^\beta_N(x)$ \eqref{rhc_pclf_2_qp} are the same, and if \ $\useq_{\beta}^0(x)$ solves 
$\mathbb{P}^\beta_N$ \eqref{rhc_pclf_2} and $(\bar{\useq}_{\beta}^0(x),\xi^0)$ solves $\bar{\mathbb{P}}^\beta_N(x)$ \eqref{rhc_pclf_2_qp}, then $\bar{\useq}_{\beta}^0(x) = \useq_{\beta}^0(x)$.
\end{rem}

\begin{prop}
The nonlinear optimization problem $\mathbb{P}^{\lambda}_N$ \eqref{rhc_pclf_1} is equivalent to the QP problem
\begin{multline} \label{eq:barPlambda}
\bar{\mathbb{P}}^{\lambda}_N(x) : \qquad \min_{\useq} V_N(x,\useq) \quad
\text{s.t.} \quad \useq \in \mc{U}^p_{N} (x), \\
F\left( A x + B u(1) \right) \leq \lambda \max(F x) \underline{1}_{r}
\end{multline}
\end{prop}
\begin{pf}
The nonlinear constraint \eqref{rhc_pclf_1} is equivalent to the linear one $F\left( A x + B u(1) \right) \leq \lambda \max(F x) \underline{1}_{r}$ in view of Lemma \ref{lem:decay}.
\qed
\end{pf}

\begin{rem}
MPC~1b can also be formulated as a QP problem \eqref{rhc_pclf_2_qp} with $\beta \doteq \beta^\star (x)$.
In MPC~1b, the price to pay for using the ``lowest but still safe'' weight $\beta^\star(x)$ \eqref{eq:beta_star} is the online computation of $\lambda^\star(x)$ from \eqref{eq:lambda-test}, that could be even more demanding than the QP \eqref{rhc_pclf_2_qp} itself. Nonetheless, problem \eqref{eq:lambda-test} can be solved off-line for a finite number of polyhedral level sets of $V_p(\cdot)$, thus providing a look-up table for $\lambda^\star(x)$, and so for $\beta^\star(x)$, depending on the ``polyhedral annulus'' the current $x$ belongs to.
\end{rem}

\section{Application examples} \label{sec:simulation}
While we proved that the proposed MPC algorithms guarantee ES on the maximal controlled domain of attraction, in the examples presented in this section we heuristically show that they can also lead to ``good'' closed-loop performances, even with short prediction horizons.

In the examples, the maximal controlled set $\mc{X}_{\infty}$ is numerically computed according to \cite{miani:savorgnan:2005}, with a tolerance $\epsilon$ and so with guaranteed contraction $\lambda = 1-\epsilon$. 
The set $\bbX_f$ is chosen as in \eqref{eq:terminal_set}, and in the computations of the DoA 
for conventional QCLF-based MPC, $\mc{X}_N$, the terminal set $\bbX_f$ is approximated with a polytope of 1000 vertices.
Also the set $\tilde{\mc{X}}_N$ \eqref{eq:tildeXN} is computed numerically for comparison, although it is not required for implementation of MPC~1a.
In both examples, the prediction horizon of the presented MPC algorithms is fixed to $N=2$, and the cost  matrices are chosen as $Q = I$ and $R = 0.1 I$.

\subsection{Example~1}
The first example is the open-loop-unstable system
\begin{equation*}
x^+ = \left[
\begin{array}{cc}
1.1 & 0 \\
0.2 & 1.1
\end{array}
\right]x + \left[
\begin{array}{cc}
0.1 & 0.1 \\
0.1 & 0
\end{array}
\right] u .
\end{equation*}
The constraint sets are
\begin{multline*}
\bbX = \{x\in \bbR^2 \mid \left\| x \right\|_\infty \leq 1 \}, \\
\bbU = \{(u_1,u_2) \in \bbR^2 \mid u_1 \in [0,1], \ u_2 \in [-1,1]  \}.
\end{multline*}
The set of admissible controls $\bbU$ has the origin in its boundary, because of the constraint $u_1 \geq 0$. As a consequence, the set $\bbX_f$ associated to the Riccati-optimal quadratic shape reduces to origin.
In order to avoid this, see \cite{rao:rawlings:1999,pannocchia:wright:rawlings:2003}, $\tilde{P}$ is chosen as the solution of the ARE associated to the matrices $A$, $B_2$ (the second column of $B$), $Q$ and $R_{2,2}$ (element in position (2,2) of matrix $R$). 
Then, $\bbX_f \doteq \{x \in \bbR^2 \mid x^\top \tilde{P} x \leq \alpha\}$ is chosen as the largest set of the kind made controlled-invariant by $u(x) = \tilde{K} x$, with $\tilde{K} = - \left( B_2^\top \tilde{P} B_2 + R_{2,2} \right)^{-1} B_2^\top \tilde{P} A$.

\begin{figure}[tb]
\begin{center}
\includegraphics[width = 0.95 \columnwidth]{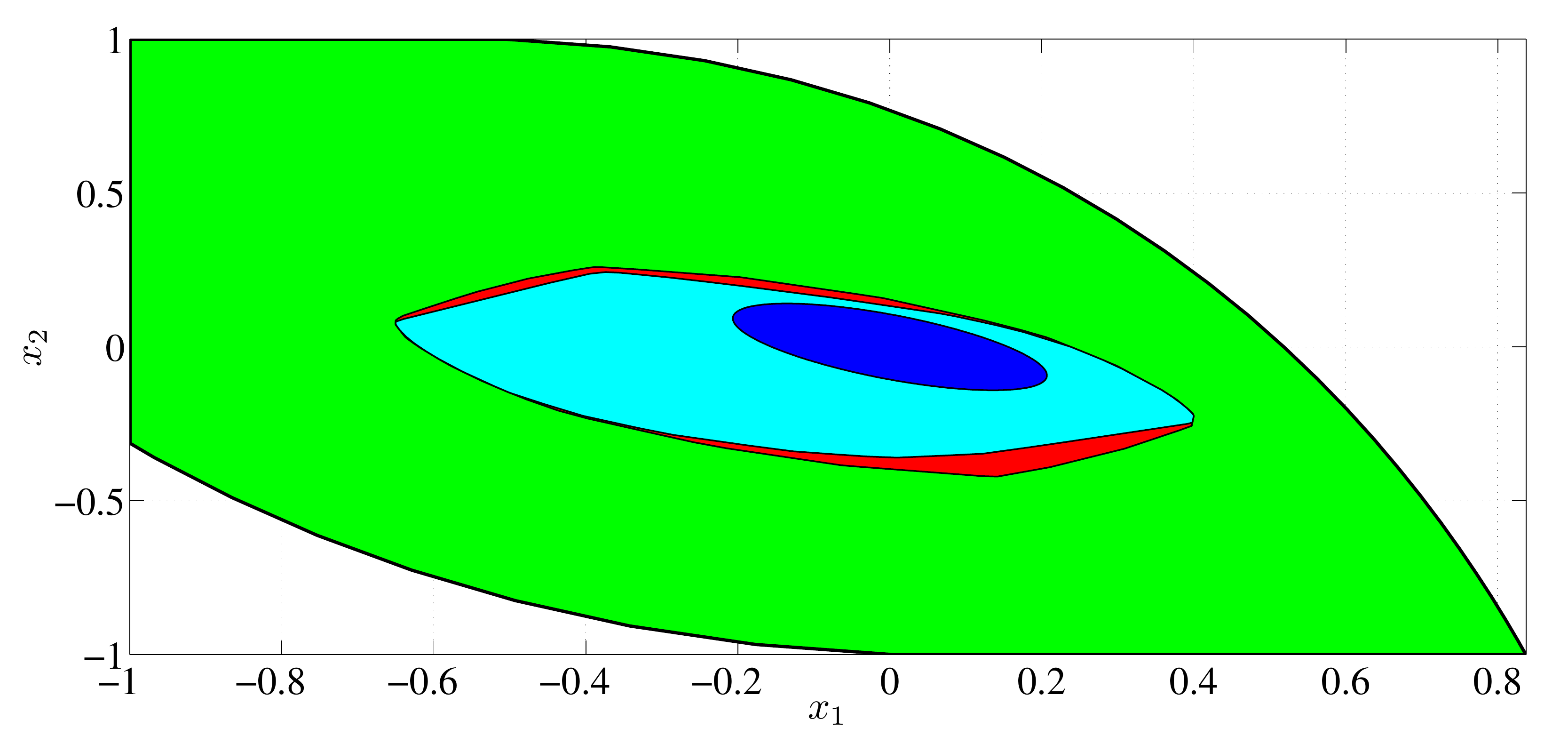}
\end{center}
\caption{Example 1: sets $\bbX_f$ (blue), $\tilde{\mc{X}}_N$ (cyan), $\mc{X}_N$ (red), $\mc{X}_{\infty}$ (green).}
\label{fig:ex2}
\end{figure}

We numerically compute the set $\mc{X}_\infty$ with a tolerance $\epsilon = 10^{-3}$. Notice that the asymmetry of $\bbU$ induces sets $\tilde{\mc{X}}_N$, $\mc{X}_N$, $\mc{X}_\infty$ to be asymmetric, as shown in Figure \ref{fig:ex2}.
Closed-loop simulations are performed for 120 steps. The results for the closed-loop performance cost, averaged over 20 simulations with initial state close to the boundary of $\mc{X}_\infty$, are shown in Table \ref{tab:exs}. 
In MPC~1 and MPC~1a, $\beta$ is chosen according to Lemma~\ref{lem:beta}, with
$\lmax_Q=1$, $c=1$, $\lmax_R = 0.1$, $\alpha_3 = (1-\lambda^2)\alpha_1 = (1-(1-10^{-3})^2) 0.7$, namely $\beta = \beta^* = 786.1$. In MPC~1b, $\beta^\star(x)$ varies from $\beta^* = 786.1$ for $x$ close to the boundary of $\mathcal{X}_\infty$ to about $1.6$ for $x$ close to the origin.
We note that the proposed MPC algorithms lead a performance that is only $3-4\%$ worse than the optimal one (resulting from solving $\mathbb{P}_N$ \eqref{standard_rhc} with horizon $N=120$). 

\begin{table}
\caption{Performance cost of the proposed MPC algorithms, normalized with respect to the optimal one. Results are obtained averaging over 20 simulations starting from initial conditions taken close to the boundary of $\mc{X}_{\infty}$.}
\label{tab:exs}
\centering
\begin{tabular}{@{}ccccccc@{}}\toprule
 & MPC~1 & MPC~1a & MPC~1b & MPC~2  \\\midrule
\textsc{Example 1}  & 1.035 & 1.026  & 1.024 & 1.034 \\
\textsc{Example 2}  & 1.065 & 1.058  & 1.004 & 1.031   \\
\bottomrule
\end{tabular}
\end{table}

\subsection{Example~2}
Consider the open-loop-unstable system
\begin{equation*}
x^+ = \left[
\begin{array}{cc}
1.0250 & 0.0125 \\
0.0250 & 1.0500
\end{array}
\right]x + \left[
\begin{array}{cc}
0.05 & 0 \\
0 & 0.05
\end{array}
\right] u ,
\end{equation*}
which is a discrete-time counterpart of the system simulated in \cite[Sec. 4]{elfarra:mhaskar:christofides:2004}.
The constraint sets are 
\begin{equation*}
\bbX = \{x\in \bbR^2 \mid \left\| x \right\|_\infty \leq 1 \}, \quad \bbU = \{u \in \bbR^2 \mid \left\| u \right\|_\infty \leq 1 \}.
\end{equation*}

\begin{figure}
\begin{center}
\includegraphics[width = 0.95 \columnwidth]{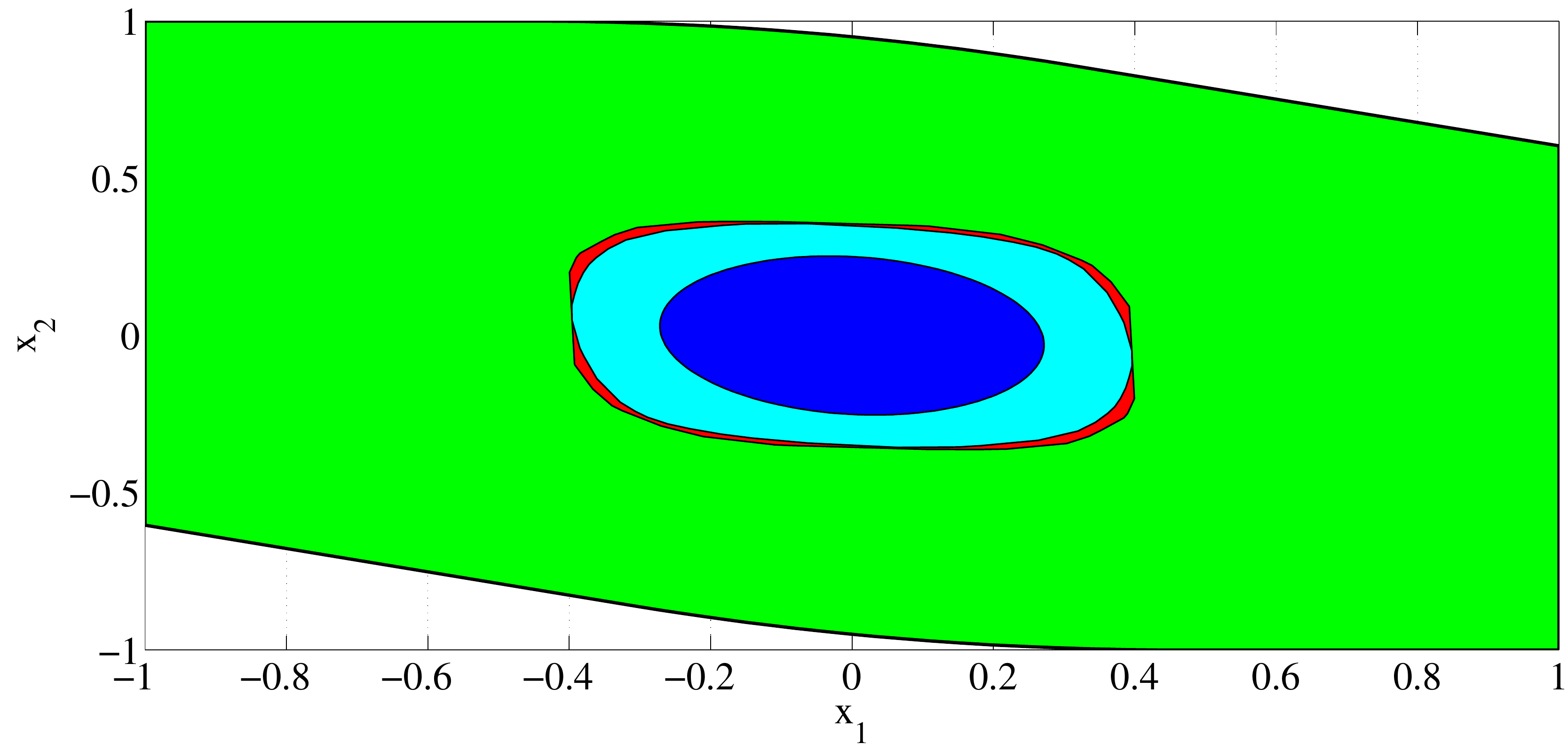}
\end{center}
\caption{Example 2: sets $\bbX_f$ (blue), $\tilde{\mc{X}}_N$ (cyan), $\mc{X}_N$ (red), $\mc{X}_{\infty}$ (green).}
\label{fig:ex3}
\end{figure}

The set $\mc{X}_\infty$, shown in Figure \ref{fig:ex3} along with $\bbX_f$, $\tilde{\mc{X}}_N$ and $\mc{X}_N$, is computed with a tolerance $\epsilon = 10^{-2}$. Closed-loop simulations are performed for 100 steps. The results (again averaged over 20 simulations with initial state close to the boundary of $\mc{X}_\infty$) for the closed-loop performance cost are shown in Table \ref{tab:exs}.
In MPC~1 and MPC~1a, $\beta$ is chosen according to Lemma~\ref{lem:beta}, with $\lmax_Q=1$, $c=1$, $\lmax_R = 0.1$, $\alpha_3 = (1-\lambda^2)\alpha_1 = (1-(1-10^{-2})^2) 0.45$, namely $\beta = \beta^* = 122.8$. 
In MPC~1b, $\beta^\star(x)$ goes from $\beta^* = 122.8$ for $x$ close to the boundary, to about 
$2.4$ for $x$ close to the origin.
All algorithms, using $N=2$, lead to 
a performance that is at most $5-6\%$ worse than the optimal one (resulting from solving $\mathbb{P}_N$ \eqref{standard_rhc} with horizon $N=100$). We also notice that MPC 1b is only $0.4\%$ worse than the optimal control.

\subsection{Further discussion on numerical results}
As expected, MPC~1a induces a better closed-loop performance with respect to the one of MPC~1. In fact, unlike MPC~1, in the set $\tilde{\mathcal{X}}_N$ MPC 1a switches to the optimal control, because the prediction $x_N \in \bbX_f$ is obtained without imposing a terminal constraint.
On the other hand, from our numerical experience, if $\lambda$ is quite close to $1$, then the constraint $\max (F \phi(1;x,\useq) ) \leq \lambda \max (F x )$ in the optimization problem $\mathbb{P}_N^\lambda$ \eqref{rhc_pclf_1}
is usually not active for $x$ close to the origin. In such cases, problems $\mathbb{P}_N^{\lambda}(x)$ \eqref{rhc_pclf_1} and $\tilde{\mathbb{P}}_N(x)$ \eqref{tilde_rhc} have the same solution. As a consequence, MPC~2 (and also its dual-mode counterpart) would show similar performance.
Finally, as expected, the closed-loop performances of MPC~1 is always improved by MPC~1b because $\beta^\star(x) \leq \beta = \beta^*$.
As a matter of fact, MPC~1b led to the best performance in all examples we tested.

\section{Conclusions} \label{sec:conclusions}
Polyhedral control Lyapunov functions (PCLFs) can shape the maximal (robust) domain of attraction (DoA) of constrained linear systems. We exploit such property to propose novel MPC formulations that guarantee the maximal controllable set, independently of the chosen finite horizon.
This result is achieved as the terminal constraint region is set to be equal to the maximal controllable set, which can be parameterized as the sub-level set of a suitably defined PCLF.
Closed-loop exponential stability of the origin is ensured either by a suitably ``inflated'' PCLF-based terminal penalty or by adding a one-step-ahead contraction constraint of the PCLF shaping the maximal DoA.
Two variants were proposed, one based on a dual-mode formulation and one based on a state-dependent terminal weight.
Moreover, the infinite-horizon optimal cost is achieved for a well defined subset of the maximal DoA, and inherent robustness with respect to arbitrary sufficiently small perturbations is proved.

Achieving the maximal DoA irrespectively of the prediction horizon is an important goal because it allows the computational burden, related to the use of long horizons, to be separated from issues of controller's feasibility and closed-loop stability. Therefore the horizon only affects the closed-loop performance.
Numerical examples showed that the DoA of the proposed formulations is much larger with respect to that of conventional MPC formulations for short horizons, and hence a great benefit in terms of feasibility is obtained.
In these examples, the closed-loop cost obtained with the proposed formulations is quite close to that of the infinite-horizon optimal controller even if very short horizons are employed.
This can be very attractive in cases where a long horizon cannot be used due to limited computational time, for instance in fast dynamic systems.
Future work will examine closed-loop nominal and robust performance from a theoretical point of view.

\bibliographystyle{elsarticle-num}
\bibliography{pclf_mpc_bib}

\end{document}